\documentclass[twocolumn,showpacs,preprintnumbers,amsmath,amssymb]{revtex4}

\usepackage{epsfig}
\usepackage{graphicx}
\usepackage{dcolumn}

\newcommand{\beq}{\begin{equation}}
\newcommand{\eeq}{\end{equation}}
\newcommand{\beqa}{\begin{eqnarray}}
\newcommand{\eeqa}{\end{eqnarray}}
\newcommand{\lam}{\lambda}

\newcommand{\ga}{\gamma}

\newcommand{\si}{\sigma}

\newcommand{\om}{\omega}

\newcommand{\ra}{\rangle}

\def\pra#1{{ Phys.\ Rev. A\/} {\bf#1}}
\def\prb#1{{ Phys.\ Rev. B\/} {\bf#1}}

\def\prl#1{{ Phys.\ Rev.\ Lett.} {\bf#1}}

\begin{document}

\draft

\title{Entanglement Decay  Versus  Energy  Change:  A Model}

\author{Ting Yu}
\email{ting@pas.rochester.edu}
\affiliation{ Rochester Theory Center for Optical Science and
Engineering, and  Department of Physics and Astronomy,
University of Rochester,  New York 14627, USA }


\date{August 28, 2006}

\begin{abstract}
We present a simple quantum open system to show quantitatively 
how entanglement decoherence is related to the energy  transfer 
between the system of interest  and its environment.  
Particularly,  in the case of the  exact entanglement decoherence  
of two qubits, we find an upper bound for the energy transfer between the  
two-qubit  system and its environments.
\end{abstract}

\pacs{03.65.Yz, 03.65.Ud, 42.50.Lc}

\maketitle

\section{Introduction}
Quantum entanglement is a central concept in the very foundation 
of quantum mechanics and recently has been
regarded as an important resource in the emerging technologies such as 
quantum information processing and quantum computing
\cite{mc,shor,glo}.    However, real quantum systems are inevitably 
influenced by their surrounding environments.  These unavoidable 
mutual interactions often result in the dissipative evolution of  quantum coherence
and loss of the useful entanglement. One of
the prominent examples of such a disentanglement process arises in
quantum registers where a set of qubits are coupled individually
to their thermal environments.

It is desirable on both fundamental and practical grounds to
understand entanglement decay in a time-dependent sense
\cite{diosi,dhal, ye2,ye3,ye4,pri,dav}.  More fundamentally,  decoherence processes due to 
the interaction with internal or external noises have been studied theoretically
in many distinct cases \cite{Zurek, Zeh,Hu-Paz-Zhang}.  
 However, it seems that a quantitative link between the degree of disentanglement 
and the amount of the energy transferred  between the system  of interest and its environment 
is still missing \cite{privman1,privman2}.

Decoherence may arise in many different situations whenever the system
of interest is not completely isolated from its environments.  Typically, the examples in
quantum optics include the so-called amplitude decoherence caused by spontaneous emissions of atoms  and  the phase decay by random disturbances of the relative phases of a quantum
state. In the latter case there is no energy exchange between the system and its environment.  
In this Letter, our focus is on the decoherence processes caused by energy dissipations 
and fluctuations.  We show explicitly how the energy  transfer is related to
a temporal disentanglement process  of mixed states. Specifically, we will address a prominent question: How can energy transfer between the system and the environment be an indicator of
the complete decoherence of quantum entanglement shared by two
qubits.  In order to analytically discuss this issue,  it is only possibly
to employ a simple model.  For our purpose, two-qubit models are ideal subjects 
as they embody in a compact way several mysterious elements  of pure or mixed states
entanglement such as Bell states and Werner states.

The Letter is organized as follows:  In Sec. \ref{model}, we present
a two-qubit model and its solutions at finite temperature. The relation between 
the disentanglement process and the energy  transfer is discussed in several interesting 
examples in Sec. \ref{energy0}. We conclude in Sec. \ref{conc}.

\section{Two-qubit Model}
\label{model}
Let us now consider a pair of  qubits $A$ and $B$ that  are
initially in an entangled state, and we assume that the two qubits
are affected by two noisy environments individually. To examine non-local
relaxation we start with a generic quantum open system model. We
let the two qubit systems interact with their local environments
${\mathcal E}_A$ and ${\mathcal E}_B$, which are modelled by very
broadband and independent sets of harmonic oscillators. We can
think of them as modes of photons or phonons. Alternatively one can
say that a single environmental reservoir for qubits $A$ and $B$
is sufficiently macroscopic that its interaction with $A$ produces
no effect on $B$, and vice versa. This is an important stipulation
because it disallows multi-qubit decoherence-free subspaces,
providing a good match with basic reality whenever spatially
well-distributed qubits are utilized.

The total Hamiltonian for the systems of interest and their
finite temperature environments can be formally written as follows ($\hbar=1$):
\beq
H_{\rm tot} = H_{\rm sys} + H_{\rm int} + H_{\rm env},
\eeq
 where
\beq H_{\rm sys} = \frac{1}{2}E_A \si^A_z+\frac{1}{2}E_B
\si^B_z\ \quad{\rm and}   \eeq \beq H_{\rm env } =
\sum_{\lambda}\omega_{\lambda} a_{\lambda}^\dag
a_{\lambda}+\sum_{\lambda}\nu_{\lambda}b_{\lambda}^\dag
b_{\lambda},\eeq for the Hamiltonians of the qubit
system and environments, and
\beq
\label{int} 
H_{\rm int} = \sum_{\lambda} ( f^*_{\lambda}\sigma_A a_{\lam}^\dagger +
f_{\lambda} \sigma_A^\dagger a_{\lambda}) + \sum_{{\lambda}}
(g^*_{\lambda}\sigma_B b^\dag_{\lambda} + g_{\lambda}\sigma_B^\dagger
b_{\lambda})\eeq for the qubit-environment interactions.
Here $\sigma_A$ and $\sigma_B$ are the system
operators coupling each qubit to its local environment, where
$f_{\lambda}$ and $g_{\lambda}$ are coupling constants and $\si_z$
denotes the usual diagonal Pauli matrix. The standard 2-qubit
basis is used: $|++\ra_,|+-\ra,|-+\ra, |--\ra,$ where $|\pm\pm\ra
\equiv |\pm\ra_A\otimes |\pm\ra_B$ denote the eigenstates of the
product Pauli matrix $\sigma_z^A \otimes \si_z^B$ with eigenvalues
$\pm 1$.  We assume that each environment is in a thermal  state  with arbitrary
finite temperature $T$.

Note that in the interaction Hamiltonian (\ref{int}) the rotating wave approximation (RWA) is used, this ensures  that the bare energy is conserved and the virtual quanta due to the presence of
the counter-rotating terms do not play a role in our discussions.  Besides, the approximation 
greatly simplifies the calculations.  It should be noted that we have assumed that coupling 
between the atoms and photon fields is so weak such that the Markov approximation
is justified.

Under usual Markov conditions the master equation for the two two-state particles
(two qubits) can be derived, and moreover the solution of the  master equation can be
expressed in terms of  $16$  Kraus operators \cite{unpublished}:
\beq
\label{kraus}
\rho(t)=\sum_{i} M_i(t)\rho(0) M_i^\dag(t),
\eeq
where $M_i$ are the Kraus operators which completely describe the reduced dynamics of
the qubits  system interacting with the finite temperature heat baths.

We show in what follows that for a class of  simple,  yet important initial states it is possible to determine analytically  the relation between the exact entanglement decoherence and the mean energy (or qubit inversion) of either qubit $A$ or qubit $B$.   To be specific, we consider all
entangled states in the very important ``standard" class \cite{ye1} of bipartite density matrices:
\begin{equation}
\label{sol} \rho(t) = \left[
\begin{array}{clcr}
a & 0  &  0 & 0 \\
0  & b & z & 0 \\
0  & z & c & 0\\
0  &  0 & 0 & d
\end{array} \right].
\end{equation}
This standard class includes the well-known pure Bell states and important mixed Werner states \cite{werner,ish}. 
Such density matrices appear naturally, for example, in spin chain models \cite{spin1,spin2} and have been used in 
various examinations of measures of entanglement \cite{standard3,standard4}. Their significance is enhanced by the fact 
that their form is preserved under noisy decoherence evolution. That is, after an interval of interaction with a noise reservoir the only 
change in $\rho$ is that the non-zero elements evolve from $a,\ b,\ c,\ d$ and $z$, which then serve as initial values. 

Under the thermal noises these initial values evolve as follows (for the zero temperature case see \cite{ye4}):
\beqa
a(t)&=&N_1\ga^4a +N_2[a+\om^2 (b +  c) + \om^4d]\nonumber\\
&&+N_3[2\ga^2a +
\ga^2\om^2( b+c)],\\
b(t)&=&N_1(\ga^2b +\ga^2\om^2 a)+N_2(\ga^2b +\ga^2\om^2 d)\nonumber\\
&&+N_3[b+\ga^4 b+\om^2( a+d) + \om^4c],\\
c(t)&=&N_1(\ga^2c +\ga^2\om^2 a)+N_2(\ga^2c +\ga^2\om^2 d)\nonumber\\
&&+N_3[c+\om^2(d + a) + \om^4 b+\ga^4 c],\\
d(t)&=&N_1[d+\om^2( b +  c) + \om^4 a]+N_2\ga^4 d\nonumber\\
&&+N_3[2\ga^2 d +
\ga^2\om^2( b+c)],\\
z(t)&=&\ga^2 z, \eeqa where the numerical factors are given by $
N_1=\frac{(\bar{n}+1)^2}{(2\bar{n}+1)^2},
N_2=\frac{\bar{n}^2}{(2\bar{n}+1)^2}$ and $
N_3=\frac{\bar{n}(\bar{n}+1)}{(2\bar{n}+1)^2}.$  Note that the
above solution was obtained under the assumption that the two
qubits are affected by the two identical local  environmental
noises, so we have $ \gamma\equiv
\gamma(t)=\exp\left[-\frac{1}{2}\Gamma(2\bar n + 1)t\right],
\om\equiv \om(t)=\sqrt{1-\gamma^2},\label{cor2}$ where  $\bar{n}$
is the mean number of quanta in the thermal
 reservoirs ${\mathcal E}_A$ and ${\mathcal E}_B$ and
$\Gamma$ is the damping rate of the qubits  $A$  and $B$.

\section{Energy transfer and exact disentanglement}
\label{energy0}

Despite considerable efforts made in the last decade, the efficient evaluations of entanglement are still restricted to a few  classes of mixed states.  In the case of two qubits, it turns out that $C(\rho(t))$, the Wootters concurrence  \cite{woo}, is a satisfactory (and conventional) tool for measuring the quantum entanglement, where $1 \ge C \ge 0$ gives the range
between the limits of maximally entangled and fully separable,
respectively. For the standard density matrix (\ref{sol}), the concurrence of $\rho(t)$ is given by \beq \label{AmplConc}
C(\rho(t))=2\max\{0, |z(t)|-\sqrt{a(t)d(t)}\}. \eeq
 We see from the above expression that the entangled  state (\ref{sol}) may  become disentangled
completely ($C \to 0$) in  a finite time. In fact, this occurs if 
\beq \label{ftime} 
|z(t)| \leq \sqrt{a(t)d(t)}.
\eeq
Now it is easy to see that when $t\rightarrow \infty$, we have
$a(t)\rightarrow N_2,  d(t)\rightarrow N_1$, but $z(t)\rightarrow 0$. Therefore, the inequality  (\ref{ftime}) is
easily satisfied for a finite $t$. By this way, we have proven that all entangled states with standard matrix
form (\ref{sol}) will completely disentangle in finite times if exposed to finite-temperature reservoirs ($\bar{n}\neq 0$). This
 is true even though, as one can easily check, the qubits $A$ and $B$  only decay exponentially to zero over an infinite time.

This simplified model allows us to determine concurrence by computing the energy change of a single qubit. For example, let us consider the energy transfer between qubit A and its environment. Here we define the energy
dissipation of  the qubit  $A$ as:  $\Delta E(t)=E(0)-E(t)$ with $E(t)=Tr(\rho H_A)$ where $H_A=E_A\sigma_z/2$. The same 
is true for qubit $B$.  For simplicity, in this paper we will only deal with  the density matrix (\ref{sol}),  the mean energy 
of qubit $A$ at $t$ is then simply given by
\beq
E(t)=\frac{1}{2}E_A(\alpha + \beta\om^2),
\eeq
where the coefficients are $\alpha \equiv a+b-c-d$ and $\beta \equiv  2(c+d-N_1-N_3)$.  Thus the $\om$ function
can be expressed in terms of the net local energy dissipation $\Delta E(t)$
during the time period $t$:
\beq
\label{energy}
\om^2=-\frac{\Delta E(t)}{E_A(c+d-N_1-N_3)}.
\eeq
By inserting (\ref{energy}) into (\ref{AmplConc}), we see that entanglement evolution is entirely determined by 
the energy change  $\Delta E$.  As a consequence of (\ref{energy}), we will now show in the following how 
the energy dissipation is related to the exact entanglement decoherence.  

We start with the zero-temperature case $T=0\ (\bar{n}=0)$ and the initial density matrix with $a=b=c=z=1/3$. We have shown \cite{ye4} that 
the solution to the master equation then takes the form: $a(t)=\ga^4/3,  b(t)=c(t)=(\ga^2+\ga^2\om^2)/3, d(t)=(2\om^2+\om^4)/3$ and 
the disentanglement is complete at the time $\Gamma t = \ln\left(\frac{2 + \sqrt 2}{2}\right)$. Here we see that
(\ref{energy}) becomes  $\om^2=3\Delta E(t)/(2E_A) $. Then from (\ref{ftime}) we know that the energy dissipation needed for
complete disentanglement is
\beq
\Delta E = \frac{2}{3}(\sqrt{2}-1)E_A \approx 0.28E_A.
\eeq

As another interesting case, we now consider 
$T\neq 0\ (\bar{n}\neq 0)$, and begin with the
well-known  Bell states 
\beq \label{bellstate}
|\Psi\rangle=\frac{1}{\sqrt{2}}(|+-\rangle  \pm |-+\rangle. 
\eeq
The above discussions have shown that the Bell states become completely
disentangled in a finite time.  First, note that the density
matrix at $t$ for the initial Bell state takes the standard form but its
time-dependent matrix elements simplify as 
\beq \label{coe1}
a(t)=N_2\om^2+N_3\ga^2\om^2,
\eeq
\beq \label{coe2} b(t)=c(t)=N_1\ga^2 +N_2\ga^2+N_3(\ga^4 +1  + \om^4),\eeq
\beq
d(t)=N_1\om^2 +N_3\ga^2\om^2,
\eeq
\beq
\label{coe4} z(t)=\pm \frac{1}{2}\gamma^2. \eeq
Next we are interested in the mean energy of qubit  $A$  at $t$,
which  is given by \beq \Delta E=-\frac{E_A\om^2}{2(2\bar n +1)}.
\eeq

Now we want to know how much energy transfer is needed in order to
completely decohere the Bell states.  For this purpose, let us
look at the concurrence (\ref{AmplConc}). Clearly,  if
$\gamma^2\leq 2\om^2\sqrt{N_1N_2}$, then (\ref{ftime}) will be
satisfied. This immediately leads to the condition, \beq \label{upperbound} 
\om^2\geq
\frac{1}{1+2N_3}=\frac{(2\bar{n}+1)^2}{(2\bar{n}+1)^2+2\bar{n}(\bar{n}+1)}
\eeq for the Bell states to be completely disentangled, i.e., for
$C(\rho)\equiv 0$. This amounts to saying that if the energy
transfer between the qubit A and its environment is more than
$|\Delta E|_{\rm ub}$ \beq \label{energy1} |\Delta E|_{\rm ub}
=\frac{E_A(2\bar{n}+1)}{2[(2\bar{n}+1)^2+2\bar{n}(\bar{n}+1)]},
 \eeq
then the Bell states (\ref{bellstate}) are completely disentangled.

More generally for $\bar{n}\neq 0$, it can be proved that
\beq \label{upperbound1} 
|\Delta E|'_{\rm ub}=\frac{|zE_A(c+d-N_1-N_2)|}{|z|+(b+c)N_3}
\eeq
provides an upper bound for all the standard matrices (\ref{sol}).  Namely, if
the energy transfer $\Delta E$ between the system and the
environment is more than $|\Delta E|'_{\rm ub}$, then all the
standard density matrices (\ref{sol}) become completely
disentangled.

\section{Conclusion}
\label{conc}
In summary, in this Letter,  for a class of simple initial entangled states, we are able to analytically 
compute the amount of energy change that corresponds  to the complete disentanglement 
of the two-qubit system. In the case of zero-temperature heat bath, the energy dissipation is 
the only source of decoherence. Our results quantitatively establish
a relation between the degree of disentanglement and the energy
dissipation. In the finite temperature case, we have provided an upper bound of the energy 
transfer that indicates the exact entanglement decoherence.  The present  paper only 
deals with the formal relation between the degree of entanglement and the energy change.
The explicit exploration of  the relation between energy changes and the degrees of disentanglement in the context of quantum measurement theory is certainly a very interesting 
issue \cite{privman1,privman2}.  Also, it should be noted that our results are based on a master equation in the Markov regimes,   it is of interest to study a more general case where non-Markovian features can be incorporated.  All of these issues will be the topics of future publications.

\section*{Acknowledgments}

We gratefully acknowledge useful discussions
with  J.H. Eberly and  Jeff Pratt.  This research was initiated under a research 
grant for which we express  appreciation to L. J. Wang and the NEC Research Institute.


\begin{thebibliography}{99}

\bibitem{mc} M.\ A.\ Nielsen and I.\ L.\ Chuang, {\it Quantum
Computation and Quantum Information} (Cambridge Univ. Press, 2000).

\bibitem{shor}  P. Shor,  SIAM  J.  Computing,  {\bf 26}, 1484(1997).

\bibitem{glo} L. \  K. \ Grover,  \prl{78},  325(1997).

\bibitem{diosi} L.\  Diosi, in {\em Irreversible Quantum Dynamics}, edited 
by F. Benatti and R. Floreanini (Springer, New York, 2003), pp. 157-163.

\bibitem{dhal} P.\ J. Dodd and J.\ J.\ Halliwell,  \pra {69}, 052105 (2004); P.\ J.\ Dodd, \pra {69}, 052106 (2004).

\bibitem{ye2} T.\ Yu and J. H. Eberly, \prb {66}, 193306 (2002).

\bibitem{ye3}T.\ Yu and J. H. Eberly, \prb {68}, 165322 (2003).

\bibitem{ye4} T.\ Yu and J. H. Eberly, \prl {93}, 140404 (2004).

\bibitem{pri} D.\ Tolkunov, V.\ Privman and P. K.\ Arvind,  \pra {71}, 060308(R) (2005).

\bibitem{dav}M. F. Santos, P. Milman, L. Davidovich, and N. Zagury,  \pra{73}, 040305(R) (2006).  

\bibitem{Zurek} W.\  H.\  Zurek, Phys. Today {\bf 44}, 1036 (1991); Prog. Theor. Phys.  {\bf 89}, 282 (1993).

\bibitem{Zeh}E.\ Joos and H.\ D.\ Zeh, Z.\ Phys.\ B {\bf 59}, 223 (1985).

\bibitem{Hu-Paz-Zhang} B. L. Hu, J. P. Paz, and Y. Zhang, Phys. Rev. D {\bf 45}, 2843 (1992).

\bibitem{privman1} An earlier discussion on the decoherence and energy change can be 
found in  V. Privman and D. Mozyrsky,  Proc. SPIE 4047, 36-47 (2000). 

\bibitem{privman2}  D.  Mozyrsky, V.  Privman,  Mod.  Phys.  Lett.  B {\bf 14}, 303 (2000).

\bibitem{unpublished} T. Yu and J. H. Eberly, in preparation (2006).

\bibitem{ye1} T. \ Yu and J. \ H. \  Eberly, e-print quant-ph/0503089.

\bibitem{werner} R. \ Werner,  \pra {40}, 4277 (1989).

\bibitem{ish} T. \ Hiroshima and S.\  Ishizaka, \pra{62}, 044302
(2000).

\bibitem{spin1} J.\ S.\  Prate,  \prl {93}, 237205 (2004).

\bibitem{spin2} J.\  Wang, H.\  Batelaan, J.\  Podany and A.\  F. \ Starace, e-print  quant-ph/0503116.

\bibitem{standard3} W.J. Mungo, et al., \pra {64}, 030302
(2001).

\bibitem{standard4} N.A. Peters, T.C. Wei and P.G. Kwiat, \pra {70},
052309 (2004).

\bibitem{woo} W.\  K. \ Wootters, \prl {80}, 2245
(1998).

\end{thebibliography}
\end{document}